\documentclass[twocolumn]{aa}
\bibpunct{(}{)}{;}{a}{}{,}
\usepackage{subcaption}
\usepackage{hyperref}
\usepackage{CJKutf8}
\usepackage[utf8]{inputenc}
\usepackage[T1]{fontenc}
\usepackage{txfonts}
\usepackage{longtable}
\usepackage{multirow}
\usepackage{lscape}
\usepackage{tabularx}

\def\apjl{ApJ}
\def\apjs{ApJS}
\def\ao{Appl.~Opt.}

\def\aap{A\&A}

\newcommand{\CNnames}[1]{{\begin{CJK}{UTF8}{gbsn}~(#1)~\end{CJK}}}

\begin{document}

\title{Nearby open clusters with tidal features: golden sample selection and 3D structure}
\titlerunning{Tidal OCs in 3D}

\author{Ming Xu \CNnames{徐铭} \inst{1} 
\and
Xiaoting Fu \CNnames{符晓婷}\inst{2}
\and
Yang Chen \CNnames{陈洋} \inst{1}\fnmsep\thanks{Corresponding author, \email{cy@ahu.edu.cn}}
\and
Lu Li \CNnames{李璐} \inst{3}
\and
Min Fang \CNnames{房敏} \inst{2}
\and\\
He Zhao \CNnames{赵赫} \inst{4,5}
\and
Penghui Liu \CNnames{刘鹏辉} \inst{2,6}
\and
Yichang Zuo \CNnames{左义昶} \inst{1}
}

\authorrunning{Xu et al.}

\institute{
School of Physics and Optoelectronic Engineering, Anhui University, Hefei 230601, China
\and
Purple Mountain Observatory, Chinese Academy of Sciences, Yuanhua Road 10, Nanjing 210023, China
\and
Shanghai Astronomical Observatory, Chinese Academy of Sciences, 80 Nandan Road, Shanghai 200030, China
\and
Instituto de Astrofísica, Dep. de Ciencias Físicas, Facultad de Ciencias Exactas, Universidad Andres Bello, Av. Fernández Concha 700, Santiago, Chile
\and
Purple Mountain Observatory and Key Laboratory of Radio Astronomy, Chinese Academy of Sciences, Yuanhua Road 10, Nanjing 210033, China
\and
School of Astronomy and Space Science, University of Science and Technology of China, Hefei 230026, China
}
\date{Received - ; accepted -}

\hypersetup{
    linkcolor=blue,
    citecolor=blue,
    filecolor=magenta,
    urlcolor=cyan
}

\date{Received .... / Accepted .....}

\abstract{Open clusters offer unique opportunities to study stellar dynamics and evolution under the influence of their internal gravity, the Milky Way's gravitational field, and the interactions with encounters. Using the Gaia DR3 data for a catalog of open clusters within 500 parsecs that exhibit tidal features reported by the literature, we apply a novel method based on 3D principal component analysis to select a ``golden sample'' of nearby open clusters with minimal line-of-sight distortions. This approach ensures a systematic comparison of 3D and 2D structural parameters for tidally perturbed clusters. The selected golden sample includes Blanco~1, Melotte~20, Melotte~22, NGC~2632, NGC~7092, NGC~1662, Roslund~6 and Melotte~111. We analyze these clusters by fitting both 2D and 3D King Profiles to their stellar density distributions. Our results reveal systematic discrepancies: most of the golden sample clusters exhibit larger 3D tidal radii compared to their 2D counterparts, demonstrating that the 2D projection effects bias the measured cluster size. Furthermore, the 3D density profiles show stronger deviations from King profiles at the tidal radii ($\Delta \rho_{\rm 3D} > \Delta \rho_{\rm 2D}$), highlighting enhanced sensitivity to tidal disturbances. Additionally, we investigate the spatial distribution of cluster members relative to their bulk motion in the Galactic plane. We find that some clusters exhibit tidal features oriented perpendicular to their direction of motion, which can be attributed to the fact that the current surveys only detect the curved inner regions of the tidal features. In conclusion, this work offers a golden sample of nearby open clusters that are most reliable for 3D structure analysis and underscores the necessity of 3D analysis in characterizing OC morphological asymmetries, determining cluster size, and identifying tidal features.}

\maketitle

\section{Introduction}

Open clusters (OCs) are key laboratories for studying Galactic structure \citep[see e.g.][]{Dias2005, CG2021, Joshi2023}, stellar evolution \citep[see e.g.][]{m67_2018, Jian2024}, star formation history \citep[see e.g.][]{Magrini2017, Spina2022, Bragaglia2024}, and the evolution of the Milky Way \citep[see e.g.][]{Kroupa2005, Magrini2017, Fu2022}. As gravitationally bound groups of stars formed nearly simultaneously from the same molecular cloud \citep{lada2003}, OCs offer a unique opportunity to explore stellar dynamics and the influence of the Galactic environment. The dynamical evolution of OCs, driven by internal processes and gravitational interactions with the Milky Way, often leads to the formation of tidal features such as tidal tails. These features reflect the cluster mass loss and dynamical evolution within the cluster \citep[see e.g.][]{Spitzer1958ApJ, Lamers2005}. Over time, the cluster continues to lose mass until it eventually disrupts and disperses all its member stars to the Galactic field. By studying these tidal features and the intrinsic properties of the star clusters, we can gain valuable insights into the internal cluster processes (e.g., gas expulsion and mass segregation) \citep{Dinnbier2020,Liu2025ApJ,Alfonso2024,Angelo2025}, external factors (e.g. the Galactic potential) \citep{Wang2021,Kroupa2022}, and the initial conditions of the Galactic field from which these stars originated \citep{Hao2023,Casado2024}.

Identifying tidal features and members of the Galactic OCs has historically been a challenging task, particularly in the pre-$Gaia$ era, when precise astrometric data, such as stellar proper motions and parallaxes, were unavailable for a great number of stars. Prior to the advent of Gaia, the identification of tidal features and cluster members primarily relied on less precise methods, including local density analysis and color-magnitude diagrams \citep[CMDs, see e.g.][]{Dalessandro2015,Bica2019,Kharchenko2013,Bica2008,Dias2002,Kharchenko2005,Piskunov2007,Platais1998,Alessi2003,Drake2005}. These methods, while useful, were often limited in their ability to reliably distinguish true tidal features from field star contamination.

The advent of the $Gaia$ mission \citep{prusti2016gaia} has revolutionized the study of OCs. With its high-precision astrometric and photometric data, released in successive $Gaia$ data deliveries \citep{gaiadr1, gaiadr2, gaiaedr3, gaiadr3}, researchers have been able to identify numerous new OCs and their member stars \citep[e.g.][]{Cantat-Gaudin2018,CG2022Univ,Hao2022,He2023,Zhong2022,Noormohammadi2023,Perren2023,van-Groeningen2023,Qin2023}. The observational confirmation of long ago predicted OC tidal tails also becomes possible with Gaia \citep{Meingast2019,Roser2019}. Since then, extended tidal features, especially in nearby OCs, have also become a focal point of discussion \citep[see e.g.][]{Bhattacharya2022,Kos2024,Risbud2025A&A}. Among these, tidal tails of nearby OCs are even traced out to a total extent of almost 1 kpc when combined with N-body simulations \citep{Jerabkova2021}. And the detailed structure of the tidal tails have profound implications, such as for cluster age dating \citep{Dinnbier2022} and testing gravitational theory \citep{Kroupa2024ApJ94K}.

The structure parameters of OCs exhibiting tidal structures are typically determined through two-dimensional (2D) analyses. These analyses are often conducted using equatorial coordinate, right ascension (R.A.), and declination (DEC.) \citep[see e.g.][]{Hunt2023,Zhong2022,Camargo2021,Gao2019,Rui2019,Yeh2019,Hosek2015}, or the Galactic longitude ($l$) and latitude ($b$) \citep[see e.g.][]{GaoXinhua2024}. To mitigate the projection effects, particularly for nearby OCs or those at high declinations, projected 2D coordinates are also employed to characterize OCs' structure \citep[see e.g.][]{Olivares2018, Tarricq2022}.

While 2D analyses provide valuable information, 3D studies may offer a more comprehensive picture of cluster morphology and dynamics, allowing for direct comparisons with N-body simulations \citep[see e.g.][]{wang2016} and revealing the true spatial distribution of stars both within and around the outskirts of clusters. However, the 3D structure of OCs with tidal features has been less explored in the literature. A major challenge in analyzing the 3D structure of OCs is the pronounced artificial elongation of the clusters along the line of sight, which can arise from the propagation of parallax uncertainties \citep[see e.g. the discussions in][]{Bailer-Jones2015}. Parallax inversion, the simplest method for deriving distances, may introduce complications: even for cluster member stars with symmetric parallax distributions, the resulting distance distributions become asymmetric. This effect was analyzed in detail by \citet{Luri2018} using Gaia DR2 data and further investigated by \citet{Piecka2021}, who quantified the impact of parallax uncertainties on the 3D morphology of clusters. They concluded that for a fixed parallax uncertainty, the apparent elongation of a cluster becomes more severe with increasing distance. Based on these findings, they recommended excluding OCs beyond 500 pc when studying 3D structures, as these clusters are more susceptible to such distortions.

To reduce the line-of-sight distortions in the 3D structure of OCs caused by parallax uncertainties, several studies have developed and applied distance corrections for OC member stars. For instance, \citet{Carrera2019} and \citet{Pang2021ApJ} adopted the Bayesian approach suggested by \citet{Bailer-Jones2015} to derive the 3D morphology of OCs, utilizing their preferred priors. Similarly, \citet{Meingast-Stefan2021A&A} refined the distances of their OC member stars by employing a mixture of 3D Gaussians as a distance prior, combined with the Gaussian likelihood derived from Gaia parallax measurements.

Another approach to mitigate the effects of line-of-sight elongation is to carefully select a sample of OCs that are less affected by such distortions. In this work, we adopt a 3D principal component analysis (PCA) method to identify nearby OCs that exhibit minimal line-of-sight elongation effects. This approach enables us to construct a ``safe'' sample of OCs, which we use to investigate how tidal interactions shape cluster morphology, compare results from 2D and 3D modeling, and assess the influence of the Galactic potential on cluster structure and longevity.

The paper is organized as follows. In Sec.~\ref{sec:data}, we describe the data used in this work. Sec.~\ref{sec:select} describes the selection of our golden sample clusters. We analyze the tidal radius, and tidal structure in Sec.~\ref{sec:method}. The results are discussed in Sec.~\ref{sec:discu}. Finally, we summarize our results and conclusions in Sec.~\ref{sec:sum}.

\section{Data}
 \label{sec:data}

The catalog by \citet[][hereafter T22]{Tarricq2022} serves as the basis for our work. This is a catalog of OCs with tidal features based on Gaia EDR3 data. They focus on 389 OCs with heliocentric distances less than 1.5\,kpc, and ages greater than 50 Myr selected from the well-known OC list of \citet[][hereafter CG20]{Cantat-Gaudin2020}. T22 identified stars within 50 pc of the cluster centers that meet the following criteria: $G<18$ mag, re-normalized unit weight error (RUWE)$<$1.4, and proper motion within 10 $\sigma$ of the cluster's bulk value. For clusters within 500 pc, T22 applied an additional cut on the parallax to exclude stars with astrometric measurements inconsistent with the cluster's mean astrometric parameters. Cluster membership in T22 was determined using the unsupervised machine learning algorithm HDBSCAN \citep[Hierarchical Density-Based Spatial Clustering of Applications with Noise,][]{Hdbscan}. A key objective of the T22 catalog is to include stars at the edges of OCs, enabling the identification of tidal structures, which was not available in the CG20 membership lists. Following the recommendations of T22, we select stars with membership probabilities of 0.5 or higher as members of these OCs.

\section{Golden sample selection}
\label{sec:select}

In this section, we describe our methodology for selecting a subset of T22 OCs that are minimally affected by line-of-sight distortions, rendering them suitable for 3D structural studies. We refer to these OCs as our ``golden sample''. Specifically, we identify these clusters by examining the alignment of their 3D PCA principal axis with the line of sight, after applying the parallax zero-point correction to the member stars.

\subsection{Parallax zero-point correction}
\label{sec:plx}

The Gaia parallax solution has been known to exhibit biases since DR2 \citep[see e.g.][]{Chan2020}, which depend on the source's projected position \citep{Lindegren2018}, as well as its magnitude and color \citep{Zinn2019}. \citet{Lindegren2021A&A} extensively discussed the systematic offset of the Gaia EDR3 parallax. To address this bias, they employed quasars, stars in the Large Magellanic Cloud, and binaries to derive the parallax zero-point offset. The corresponding parallax zero-point offset values are calculated using the Python package \texttt{gaiadr3\_zeropoint}\footnote{\url{https://gitlab.com/icc-ub/public/gaiadr3_zeropoint}} by inputting the following parameters: $G$-band magnitude (\texttt{phot\_g\_mean\_mag}), effective wavenumber (\texttt{nu\_eff\_used\_in\_astrometry}), astrometrically estimated effective wavenumber (\texttt{pseudocolour}), five or six parameter solution (\texttt{astrometric\_params\_solved}).

For all OCs in the T22 catalog, we correct the parallaxes of their member stars. The median of these median zero-point offsets is -0.0333 mas, with the smallest offset being -0.0266$\pm$0.0114 mas and the largest being -0.0467$\pm$0.0102 mas. After correcting the zero-point offset, all T22 OCs exhibit larger parallaxes, corresponding to smaller distances. The distance standard deviation among member stars in each OC decreases slightly, indicating a marginally more compact distribution, though the variation is minimal. For OCs within\,500 pc, the maximum change in the standard deviation of the distance is -2.74 pc; for those between 500 pc and 1000 pc, it is -7.92 pc; and for those beyond 1000 pc, it is 79.08 pc. After inverting the corrected parallaxes to derive distances, the largest median distance correction within 500 pc is for UPK 24 (-11.895$\pm$3.544 pc). For clusters between 500 pc and 1000 pc, the largest correction is for NGC~6416 (-46.564$\pm$12.294 pc), and for clusters beyond 1000 pc, the largest correction is for UBC 326 (-108.056$\pm$37.153 pc). These corrections are essential for refining distance estimates and reducing the line-of-sight elongation caused by the parallax zero-point offset.

\subsection{3D PCA alignment with the line-of-sight}
\label{sec:3dpca}

After correcting the parallaxes, we perform the 3D PCA on all the T22 OCs using their Galactocentric Cartesian coordinates (X, Y, Z). The goal is to identify OCs with minimal elongation effects along the line of sight. PCA \citep[see][for a review]{pca} is a statistical method that determines the directions of the largest variance in a dataset with multi-dimensional variables. For our case with Galactocentric Cartesian coordinates (X, Y, Z) for each member star in an OC, the first principal axis of PCA captures the direction with the greatest ``apparent" spatial variance.

These Galactocentric coordinates are calculated using the Python package \texttt{astropy} \citep{Astropy-Collaboration2013}, incorporating right ascension, declination, and the zero-point corrected parallax as described earlier. To account for uncertainties in these parameters, we apply Monte Carlo sampling with 10\,000 iterations to estimate the corresponding errors in (X, Y, Z). We adopt the Solar position as [X, Y, Z]$_{\odot}$ = [-8200.0, 0.0, 14.0] pc following \cite{McMillan2017}. 

Most PCA implementations assume mean centering, making the results sensitive to outliers and scaling. However, for OCs studied here, their Galactocentric Cartesian coordinates (X, Y, Z) are expected to exhibit asymmetry and skewed distributions due to tidal structures. In such cases, median centering provides a more representative central point around which the majority of the data congregates. Therefore, we center the data by subtracting the median [X, Y, Z] values before performing PCA.

The PCA process in this work involves three key steps: 
\begin{itemize}
    \item Centering the Data. Subtract the median 3D position from the Galactocentric coordinates [X, Y, Z] of all cluster member stars to centralize the dataset. Unlike typical PCA applications, where datasets consist of multi-dimensional heterogeneous variables requiring standardization, Galactocentric coordinates are homogeneous spatial variables.
    \item Computing the First Principal Axis. Use the Python package \texttt{sklearn} \citep{scikit-learn} to compute the first principal axis of PCA. The PCA of the dataset is performed with the parameter ``n\_components" set to 1, yielding the 3D vector (m, n, p) representing the dominant direction of variance.
    \item Re-centering the Coordinates. Re-center the coordinate to express the first principal axis in Galactocentric Cartesian coordinates: $\frac{(X_{\rm P1}-X_0)}{m}=\frac{(Y_{\rm P1}-Y_0)}{n}=\frac{(Z_{\rm P1}-Z_0)}{p}$.
\end{itemize}

To assess whether an OC exhibits significant elongation along the line-of-sight, we propose using the angle between the line-of-sight vector to its center and the cluster's first principal axis derived from PCA. Specifically, the angle $\phi$ quantifies the alignment between the PCA axis and the line of sight to the cluster center. A small value of $\phi$ might indicate that the ``dominant'' spatial distribution of cluster members is strongly aligned with the line of sight, while a large $\phi$ may suggest a minimal elongation in the line-of-sight direction. We also notice that this assumption is valid only if the tidal tail is not physically along the line of sight. We will discuss this issue in more detail in the followed text.

\begin{figure}[ht] 
  \centering
  \includegraphics[width=\columnwidth]{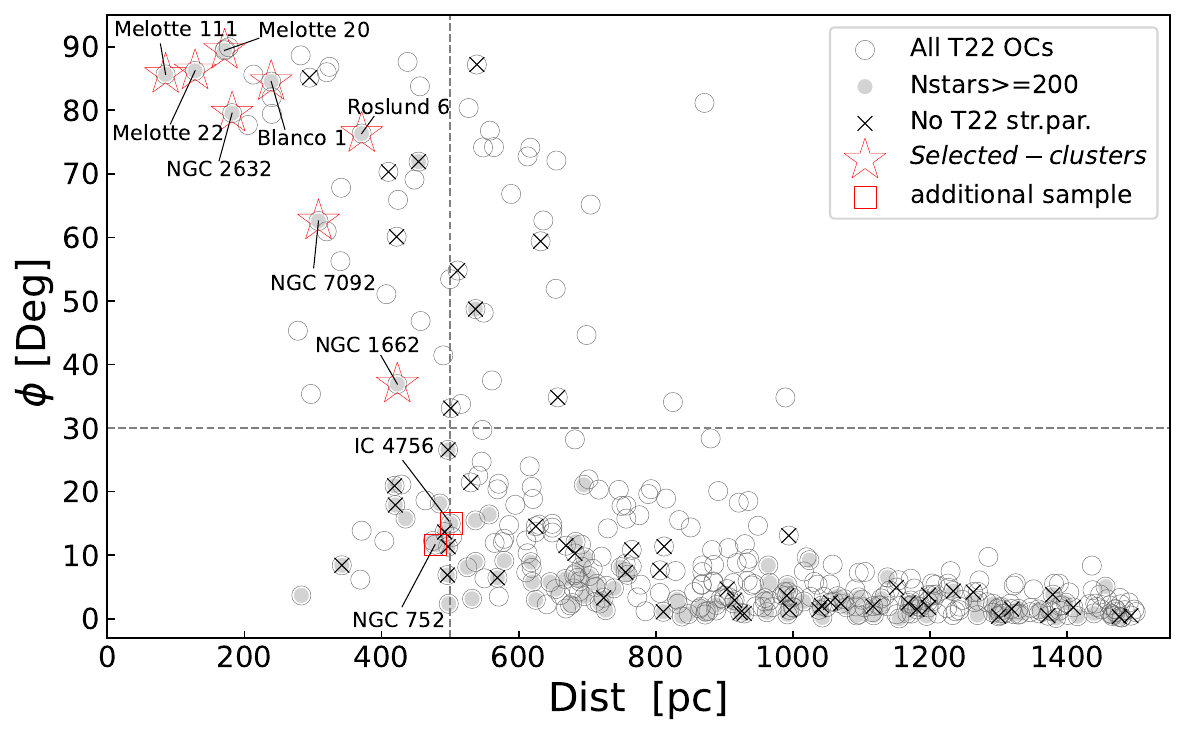} 
  \caption{The 3D angle $\phi$ between the PCA axis and the line-of-sight to the cluster center as a function of distance for all 389 OCs in the T22 catalog. OCs with more than 200 members are marked with gray filled circles, and those without structure parameters in the T22 catalog are marked with $\times$. Red empty pentagrams are the selected golden sample OCs and are labeled with the cluster names. The red empty squares indicate IC\,4756 and NGC\,752, for reference. The horizontal gray dashed line represents the $\phi= 30^\circ$ criteria and the vertical gray dashed line indicates the 500 pc distance from us.
  }
  \label{fig:dist-all}
\end{figure}

Figure~\ref{fig:dist-all} shows the distribution of the 3D angle $\phi$ as a function of heliocentric distance for all T22 OCs. In general, $\phi$ decreases with the heliocentric distance, likely suggesting a stronger effect of the line-of-sight distortion effect for distant OCs. \citet{Piecka2021} recommended excluding clusters beyond 500 pc for cluster morphology studies, a threshold also adopted by \citet{Meingast2021} and \citet{Pang2021ApJ}. Accordingly, we set an upper limit of 500 pc for our golden sample OCs. Indeed, Figure~\ref{fig:dist-all} clearly shows that $\phi$ values decrease significantly beyond $\sim$500 pc (indicated by the vertical dashed line).

To ensure statistical reliability for structural analyses, we retain only clusters with at least 200 member stars (represented by gray filled circles in Figure~\ref{fig:dist-all}). Additionally, any OCs lacking structure parameter determinations in T22 (marked with $\times$) are excluded for further analysis.

\begin{figure*}
    \centering
    \begin{subfigure}[]{\textwidth}
        \includegraphics[width=\textwidth]{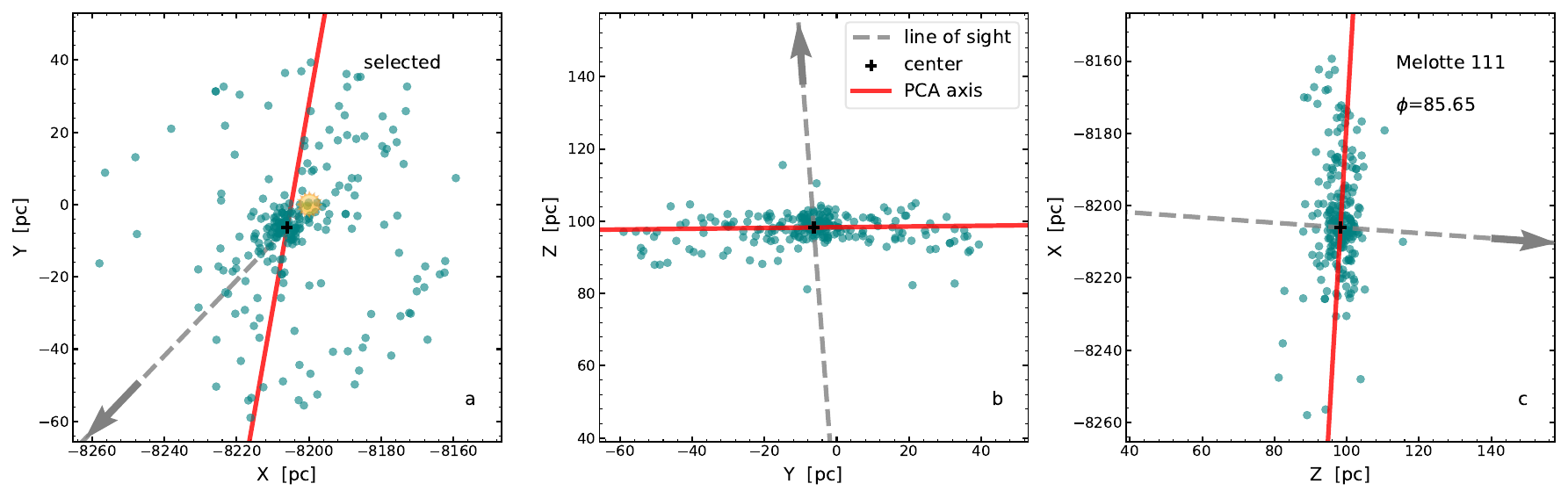}
    \end{subfigure}
    \hspace{-1cm}
    \begin{subfigure}[]{\textwidth}
        \includegraphics[width=\textwidth]{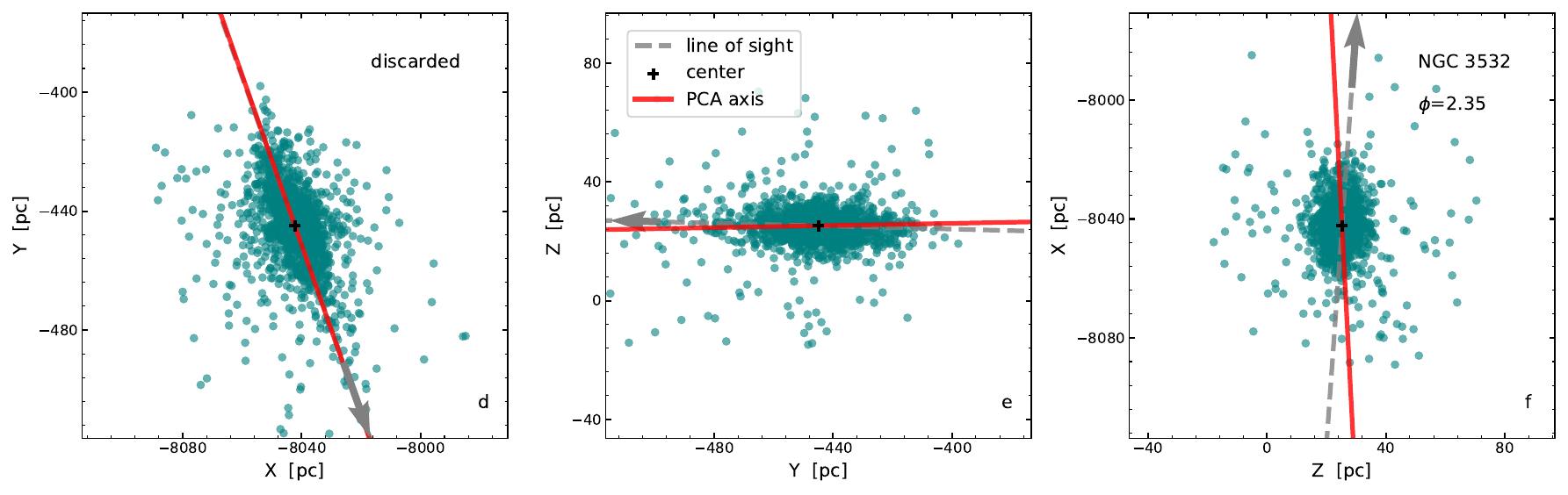}
    \end{subfigure}
    \caption{Examples of cluster morphology in Galactocentric Cartesian coordinates. The upper panel shows Melotte~111 (a high-$\phi$ cluster), while the lower panel illustrates NGC~3532 (a low-$\phi$ example). The red lines denote the projection of the first principal PCA axis in 3D, and the gray dashed lines represent the solar position to the cluster center (marked by the black cross).}
  \label{fig:xyz}
\end{figure*}

The combined selection based on distance and T22 structure parameters reveals a gap at $\phi \sim 30^{\circ}$ in the $\phi$-distance diagram. We use this gap to divide the remaining OCs into two categories: high $\phi$ and low $\phi$. Figure~\ref{fig:xyz} illustrates examples of both high $\phi$ and low $\phi$ clusters in the Galactocentric Cartesian coordinates (X, Y, Z), highlighting their PCA axis, the line-of-sight, and the 3D distribution of their member stars. Clusters in the high $\phi$ category with $\phi>30^{\circ}$ are designated as our golden sample because they exhibit minimal elongation along the line of sight. Since our goal is to define a conservative golden sample rather than a complete sample, there might be some clusters with genuine tidal structures but excluded by our selection criteria. For example, cluster IC~4756 ($\phi=15^{\circ}$) and NGC~752 ($\phi=11.66^{\circ}$), both with $\phi<30^{\circ}$, are found to present tidal structures \citep{Bhattacharya2021,Boffin2022,Ye-Xianhao2021}. They are not selected in our golden sample, but listed as additional sample clusters because of their reported tidal tails. Moreover, as mentioned before, if an open cluster has the tidal tail physically along the line of sight, it would show small $\phi$. This might cause some correlation with the Galactic longitude. However, we exclude this possibility by statistically studying the listed Galactic coordinates and $\phi$ from the full table \ref{tab:phi}.

In total, we end up with eight OCs in the golden sample, containing a total of 4374 member stars. These clusters span a wide range of ages, from 51.3 Myr to 810 Myr according to the T22 catalog. The central coordinates ($X_0$, $Y_0$, $Z_0$) of the golden sample OCs, along with their PCA first principal axis parameter (m, n, p), and the angle $\phi$, are listed in Table~\ref{tab:phi}. The corresponding parameters for all other T22 OCs are provided in the full version of the online table.

\begin{table*}[ht]
    \caption{Basic properties of the input OC sample.}
    \label{tab:phi}
    \centering
    \begin{tabular}{@{}lrrrrrrrrrrr@{}}
\hline
Cluster &$l_0$&$b_0$&X$_{0}$ & Y$_{0}$ & Z$_{0}$ & $m$ & $n$ & $p$ & $\phi$ &$N_{\rm star}$ &log\,Age\,(yr)\\
&(Deg)&(Deg)&(pc) & (pc) &(pc) & & & &(Deg) & &\\
\hline
Melotte 20 &147.40 &-6.28 & -8344.9 & 92.2 & -5.4 & -0.35 & -0.70 & -0.62 & 89.44 & 756 &7.71\\
Melotte 22 &166.58 & -23.54& -8320.5 & 29.0 & -39.7 & 0.20 & -0.71 & -0.67 & 86.21 & 1222 & 7.90\\
Melotte 111 &220.69 & 83.95& -8206.1 & -6.3 & 98.3 & 0.17 & 0.98 & 0.01 & 85.65 & 244 &8.81\\
Blanco 1   &14.41 &-78.94 & -8157.5 & 11.9 & -216.0 & 0.53 & 0.82 & 0.24 & 84.52 & 512 &8.03\\
NGC 2632   & 205.95&32.43 & -8339.2 & -67.9 & 112.5 & -0.47 & -0.54 & -0.70 & 79.56 & 812 &8.84\\
Roslund 6  &78.85 &0.57 & -8132.2 & 342.7 & 18.0 & 0.96 & 0.06 & -0.29 & 76.35 & 314 &7.96\\
NGC 7092   & 92.31& -2.44& -8211.5 & 294.6 & 2.5 & -0.77 & -0.47 & 0.43 & 62.59 & 229 &8.60\\
NGC 1662   & 187.82& -21.13& -8574.1 & -51.3 & -130.5 & -0.66 & -0.70 & -0.27 & 36.93 & 285 &8.91\\
... &&&&&&&\\
IC 4756  & 36.35 & 5.36 & -7823.9 & 276.4 & 56.6 & 0.64 & 0.77 & -0.01 & 15.00 & 510 &9.12\\
NGC 752   &136.99 & -23.29& -8492.3 & 272.9 & -157.5 & -0.80 & 0.48 & -0.36 & 11.66 & 339 &9.08\\
ASCC 58   &281.77 & 1.18& -8101.1 & -457.1 & 21.0 & 0.11 & 0.99 & -0.08 & 18.18 & 532 &7.73\\
NGC 3532  &289.56 & 1.42& -8042.3 & -444.8 & 25.2 & 0.33 & -0.94 & -0.02 & 2.37 & 1999 &8.60\\
NGC 6475  & 355.77&-4.53 & -7925.1 & -19.9 & -8.1 & 1.00 & -0.02 & -0.04 & 3.65 & 1147 &8.35\\
Stock 2   & 133.44& -1.69& -8453.9 & 267.8 & 4.1 & 0.86 & -0.51 & -0.03 & 15.87 & 1530 &8.60\\
UBC 480   &264.94 &-10.69 & -8242.2 & -466.7 & -77.8 & 0.28 & 0.95 & 0.10 & 12.28& 470 &8.32\\
... &&&&&&&\\
\hline
\end{tabular}
    \tablefoot{Column ``Cluster" lists the OC names. Columns $l_0$ and $b_0$ list the Galatic longitudes and latitudes. Columns X$_{\rm 0}$, Y$_{\rm 0}$ and Z$_{\rm 0}$ list the Galactocentric Cartesian coordinates. Columns $m$, $n$ and $p$ list the PCA first principal axis vector. Column $\phi$ lists the angle between the PCA first principal axis and the line-of-sight to the cluster center. The full table of all the T22 OCs is available online.}
\end{table*}

\section{Analysis with 3D data}
\label{sec:method}

Using the golden sample OCs -- those exhibit minimal elongation along the line-of-sight -- we perform a 3D structural analysis. Specifically, we fit each cluster's spatial distribution with a 3D King profile \citep{king1962} and examine its 3D morphology in relation to its Galactic motion. Before the analysis, we are aware that the stellar membership of the T22 catalogue may not be complete, so all results based on this catalog should be interpreted accordingly. One of the the major incompleteness is that the cluster members are limited to 50\,pc only, while in reality the tidal tails can extend to several hundred parsecs \citep[see e.g.][]{Jerabkova2021,Kos2024}.

\subsection{King profile fitting}
\label{sec:density}

To quantitatively characterize the structure of the selected golden sample OCs, we derive both 2D and 3D structure parameters by fitting a King profile \citep{king1962} to their respective number-density profiles. King profile is widely used in the literature for fitting the density profile of open clusters. However, we call the reader's attention that the King profile is derived assuming a spherically symmetrical mass distribution, therefore any structure deviating from this symmetry may bring errors to the fitting parameters. In this work, King profile fitting suffices our purpose of comparing the 2D and 3D structure parameters in the same manner, despite its limits.

We employ a non-linear least squares approach to estimate the best-fit parameters. In this study, we adopt the following functional form: 
\[
n(R) = 
\begin{cases}
  k (\frac{1}{\sqrt{1+(R/R_{\rm c})^2}}-\frac{1}{\sqrt{1+(R_{\rm t}/R_{\rm c})^2}})^2+c, & \text{if } R < R_{\rm t} \\
  c, & \text{if } R \geq R_{\rm t}
\end{cases}
\]
where, \(k\) is a proportionality constant related to the central density of the cluster; \(R_{\rm c}\) is the core radius; \(R_{\rm t}\) is the tidal radius; and \(n(R)\) represents either the 3D or 2D number density (in \(\mathrm{pc}^{-3}\) or \(\mathrm{pc}^{-2}\), respectively). The constant \(c\) accounts for the background level of field stars. Although our member stars are already of high probability, we follow \citet{kroupa2010} to include a background term to improve the stability and quality of the King model fits.

To construct the 3D number-density King profile, we first compute the 3D radial distance of each star relative to the cluster center using:
\[
D_{\rm C} = \sqrt{(x - X_0)^2 + (y - Y_0)^2 + (z - Z_0)^2},
\]
where \(X_0\), \(Y_0\), and \(Z_0\) are the median center coordinates listed in Table~\ref{tab:phi}. We divide each cluster into 20 radial bins: the innermost 10 bins are spaced 1 pc apart, and the remaining 10 outer bins divide the radial range from 10 pc out to the furthest member star in equal logarithmic intervals. If the bin \(i\) contains no stars, it is merged with the adjacent bin \(i+1\). We take the mean radial distance of the stars in each bin to represent the bin's distance to the cluster center. The corresponding 3D number density in a spherical shell with inner radius $D_i$ and outer radius $D_{i+1}$ is defined as:
\[
\rho_{\rm 3D} = N/V =\frac{N}{4/3 \pi \bigl(D_{i+1}^3 - D_{i}^3\bigr)},
\]
where \(N\) is the number of stars in the bin. This procedure yields a set of paired values \((D, \rho_{\rm 3D})\), which we use for the King profile fitting.

For the 2D case, we project each star's position onto a plane using the following equations \citep{vanven,Olivares2018}:
\[
\begin{aligned}
x &= D \, \sin(\alpha - \alpha_{\rm c}) \, \cos(\delta), \\
y &= D \bigl[\cos(\delta_{\rm c}) \, \sin(\delta) 
  \;-\; \sin(\delta_{\rm c}) \, \cos(\delta)\,\cos(\alpha - \alpha_{\rm c})\bigr],
\end{aligned}
\]
where \(D\) is the heliocentric distance (in pc) of the cluster according to T22, and \(\alpha_{\rm c}\) and \(\delta_{\rm c}\) are the median right ascension and declination of the cluster. The 2D radial distance for each star is computed as \(R = \sqrt{x^2 + y^2}\). The binning strategy mirrors the 3D case: we form 20 radial bins (merging any empty group with its neighbor) and take the average \(R\) within each bin. The 2D number density in a circular annulus surface is given by:
\[
\rho_{\rm 2D} = N/S = \frac{N}{\pi \bigl(R_{\rm b}^2 - R_{\rm a}^2\bigr)},
\]
where \(R_{\rm a}\) and \(R_{\rm b}\) are the inner and outer radii of the bin, respectively, and \(N\) is the number of stars in the bin. This process yields a 2D radial number-density profile \((R, \rho_{\rm 2D})\) for each cluster.

To fit the King profiles to both the 2D and 3D number-density profiles, we use the non-linear least squares method \citep[via the Python package \texttt{SciPy.optimize},][]{SciPy}. In the fitting routine, we define the residuals as ($f(x) - y)/e_y$, where \(f(x)\) is the density obtained from the King function, \(y\) is the observed density, and \(e_y\) is the Poisson uncertainty of \(y\). From these fits, we obtain the core radius \(R_{\rm c}\), tidal radius \(R_{\rm t}\), background \(c\), and proportionality constant \(k\) for each OC.

\subsection{Cluster morphology to its motion}
\label{sec:orbit}

\begin{figure}
  \centering
  \includegraphics[width=\columnwidth]{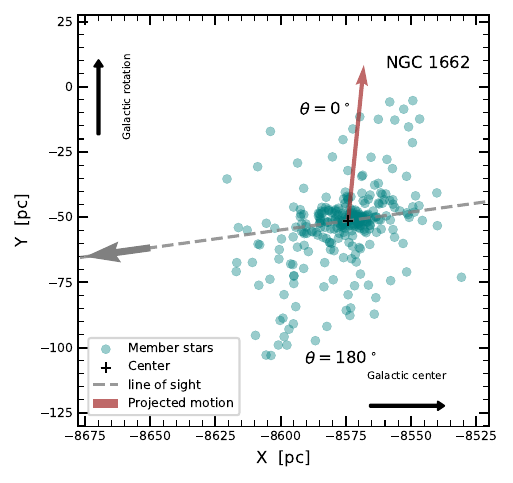}
  \caption{NGC~1662 morphology in the Galactic X-Y plane. The brown arrow indicates the future motion direction. For reference, the direction of the Galactic center and the galactic rotation are also illustrated. The line-of-sight direction to the cluster center (marked by the black cross) is also plotted as the gray dashed arrow. The position angles $\theta=0^\circ$ and $\theta=180^\circ$ represent the projected direction of the cluster's forward motion and rearward direction.}
  \label{fig:orbit}
\end{figure}

To explore the relationship between cluster morphology and the direction of motion, we use the Galactocentric Cartesian coordinates as calculated in Sec.~\ref{sec:3dpca} on the X-Y plane for further analysis. The cluster members are represented as teal circles in Figure~\ref{fig:orbit} for NGC~1662 as an example. 

We determine the bulk motion of each cluster on this plane by computing the median velocity vector (see Table~\ref{tab:uvw} for all golden sample clusters) of the member stars. The resulting vector, indicated by the gray arrow in Figure~\ref{fig:orbit}, defines the position angle $\theta=0^\circ$, which we term the forward motion of the cluster. The opposite direction, $\theta=180^\circ$, denotes the rearward direction relative to the cluster's motion through the Milky Way.

This convention ($\theta=0^\circ$ and $\theta=180^\circ$) is motivated by the well-known phenomenon of tidal structures in OCs. As an OC orbits the Galaxy, it can develop both leading and trailing tidal tails, where stars are preferentially stripped ahead of and behind the cluster along its orbit due to the differential gravitational forces in the Galaxy. By defining the cluster's instantaneous motion vector in this way, we can examine whether the spatial distribution of member stars exhibits elongations aligned with (or against) the direction of travel. In particular, an overdensity of cluster members in the region near $\theta=0^\circ$ might be interpreted as a leading tidal feature, whereas an overdensity near $\theta=180^\circ$ could indicate a trailing tidal tail.

Figure~\ref{fig:orbit} illustrates these morphology and motion considerations. The direction toward the Galactic center is shown by the horizontal black arrow, while the vertical black arrow indicates the direction of Galactic rotation. The black cross marks the adopted cluster center (see Table~\ref{tab:phi}). The gray arrow, centered on the cluster, highlights the projected direction of future cluster motion (i.e. $\theta=0^\circ$) in the X-Y plane. By combining the cluster morphology on the X-Y plane with these reference vectors, we can assess whether the distribution of stars reveals an extended structure aligned with the orbit.

\begin{table}[ht]
    \caption{The bulk motion of the golden sample OCs and the two additional samples.}
    \label{tab:uvw}
    \centering
    \begin{tabular}{@{}lrrrrrrr@{}}
\hline
Cluster &U$_{0}$ & V$_{0}$ & W$_{0}$ \\
  &(km/s) & (km/s) &(km/s) \\
\hline
Melotte 20 & -2.151 & 220.912 & 0.672\\
Melotte 22 & 4.485 & 216.731 & -6.743\\
Melotte 111 & 8.717 & 239.505 & 6.654\\
Blanco 1 & -7.339 & 238.447 & -1.926\\
NGC 2632 & -31.442 & 224.779 & -2.084\\
Roslund 6 & 0.958 & 238.255 & 1.209\\
NGC 7092 & 37.979 & 239.202 & -5.609\\
NGC 1662 & 24.744 & 245.012 & 8.292\\
\hline
IC 4756 & -2.977 & 224.138 & -2.533\\
NGC 752 & -5.331 & 224.813 & -11.427\\
\hline
    \end{tabular}
    \tablefoot{Column ``Cluster" lists the OC names. Columns U$_{0}$, V$_{0}$ and W$_{0}$ list the cluster members' median velocities in the Galactocentric Cartesian coordinate system.}
\end{table}

\section{Results and Discussions}
\label{sec:discu}

Based on our selected OC golden sample, which exhibits minimal line-of-sight distortions and is suitable for 3D structure analysis (see Sec. \ref{sec:select} for the selection process), we present the results and the corresponding discussions in this section. Specifically, we examine the tidal radius differences between 3D and 2D analysis and the position angle density distribution relative to the direction of motion. 

\subsection{Tidal radius difference in 2D and 3D}
\label{sec:2d3d}

The tidal radius of a star cluster determined with the King profile fitting represents the approximate distance from the cluster center beyond which stars are more likely to be stripped away by the gravitational field of the host galaxy than retained by the cluster's self-gravity. A star cluster with a small tidal radius loses stars more rapidly due to tidal interactions with the Galaxy. From a dynamical perspective, we can roughly consider the tidal radius as the ``edge'' of the cluster, as stars beyond this radius are not bound to the cluster anymore.

A natural question is whether the tidal radius inferred from 2D projected data differs from the 3D intrinsic value. In other words, does the choice of 2D or 3D analysis lead to different estimates of the cluster's apparent edge? To address this, we compare the King profile fitting results for our golden sample OCs in both 2D and 3D, as listed in Table~\ref{tab:structure-parameters}, with the fitting procedure described in Sec.~\ref{sec:density}.

\begin{figure}
  \centering
  \includegraphics[width=\columnwidth]{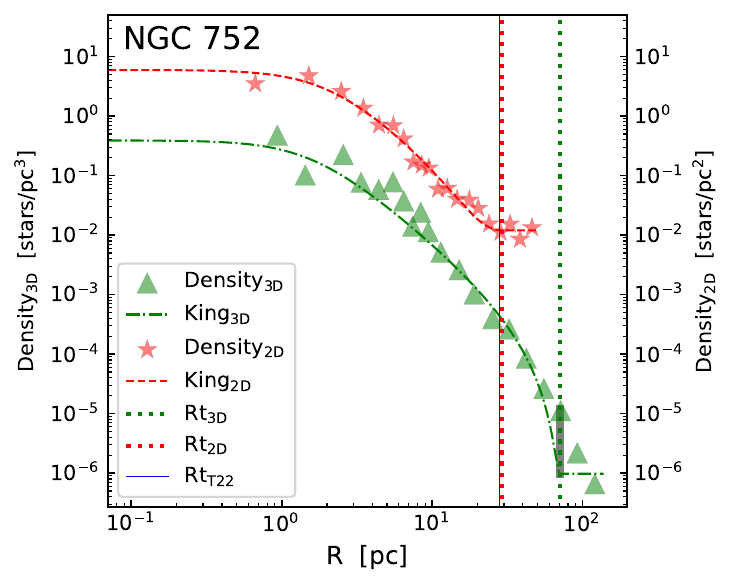}
  \caption{The 3D and 2D King profile fitting for the cluster NGC\,752. The X-axis represents the radial distance from the cluster center, while the left Y-axis corresponds to the 3D density and the right Y-axis to the 2D density. The pink stars, dashed line, and vertical dotted line indicate the observed 2D number density, the fitted King profile, and the corresponding tidal radius, respectively. Similarly, the green triangles, dot-dashed line, and vertical dotted line represent the 3D counterparts. For comparison, we also overplot the tidal radius given by T22 as a blue solid vertical line. The deviation of the data from the 3D King profile at the tidal radius is highlighted with a dark green vertical shaded region.
  }
  \label{fig:King-true}
\end{figure}

Based on the same member star catalog, four of the golden sample OCs show a larger 3D tidal radius, $R_{t,{\rm 3D}}$, compared to their 2D tidal radius, $R_{t,{\rm 2D}}$. This indicates that these clusters appear to have a larger ``edge'' when analyzed in 3D than when using the conventional 2D method. However, four clusters -- Melotte~111, Melotte~20, NGC~7092, and Roslund~6 -- show nearly the same or slightly smaller $R_{t,{\rm 3D}}$ than $R_{t,{\rm 2D}}$. Figure~\ref{fig:King-true} illustrates the 3D and 2D King profile fits for NGC~752, which has a pronounced discrepancy with $R_{t,{\rm 3D}}=71.02$ pc (green vertical dotted line) versus $R_{t,{\rm 2D}}=29.23$ pc (pink vertical dotted line), though it is in the additional sample. In the figure, the 3D data and the fitted King profile are represented by green triangles and a green dot-dashed line (with density on the left Y-axis), while the 2D data are marked with pink stars and a pink dashed line (with density on the right Y-axis). Although the King profile assumes a spherical symmetry, clusters with tidal structures can be elongated in different directions. Stars that are physically distant from the center along the line of sight may appear closer in projection, thereby biasing the density distribution inward and resulting in a smaller apparent 2D tidal radius. To test this speculation, we perform Monte Carlo simulations with 3D and 2D King profiles. We simulate star clusters and compute the corresponding 3D King profile number density $\rho(r)$ and the projected 2D surface number density profile $\mu(r)$. By doing it, we can fit 3D and 2D tidal radius with the same set of simulated data. We do find that with the King profile in spherical symmetry, the 3D tidal radii are larger than the 2D ones and the difference varies with the specific King profile parameters. However, we should not omit that the line-of-sight distortions (though minimized through our selection criteria), asymmetric tidal structures, and Poission fluctuations may have also played important roles in the observed tidal radius difference in our sample.

To quantify how significantly the observed density profile deviates from the King model in 3D versus 2D, we introduce the parameters $\Delta \rho _{\rm 3D}$ and $\Delta \rho _{\rm 2D}$, defined as
\[
  \Delta\rho=(\rho_{\rm obs}-\rho_{\rm mod})/\rho_{\rm mod} 
\]
 where $\rho_{\rm obs}$ is the observed number density and $\rho_{\rm mod}$ corresponds to the King model number density. At the tidal radius, $\rho_{\rm mod}$ equals the background level of field stars $c$ for a spherical symmetry cluster in equilibrium (see Sec.~\ref{sec:density}). A larger deviation of the observed number density from the King profile implies a more distorted cluster \citep[see discussions in ][]{Dalessandro2015}. Thus, $\Delta \rho _{\rm 3D}$ and $\Delta \rho _{\rm 2D}$ quantifies the deviation of the observed number density from the model at the tidal radius, expressed in units of the background density $c$. In Figure~\ref{fig:King-true}, $\Delta \rho _{\rm 3D}$ is shown as a dark green vertical shaded region, whereas $\Delta \rho _{\rm 2D}$ (represented by a red shaded region) is so small that it is nearly invisible. The values for all OCs in our sample are summarized in Table~\ref{tab:structure-parameters}. For every cluster, $\Delta \rho _{\rm 3D}$ is significantly larger than $\Delta \rho _{\rm 2D}$, suggesting that the 3D King profile provides a more sensitive indicator of tidal departures than the classical 2D approach.

\subsection{Density distribution in position angle}
\label{sec:morphology}

In Sec.~\ref{sec:orbit} we analyzed the cluster morphology in the Galactic X-Y plane with respect to its bulk motion for each of the golden sample OCs, aiming to characterize whether the member star distribution reveals any extended structure along the cluster's motion in the Galactic Plane. We quantify these potential tidal features by examining star-count radial profiles as a function of $\theta$, and by comparing the derived density profiles in the leading ($\theta=0^\circ$) versus trailing ($\theta=180^\circ$) regions. 

\begin{figure*}[ht]
  \centering
  \includegraphics[width=\textwidth]{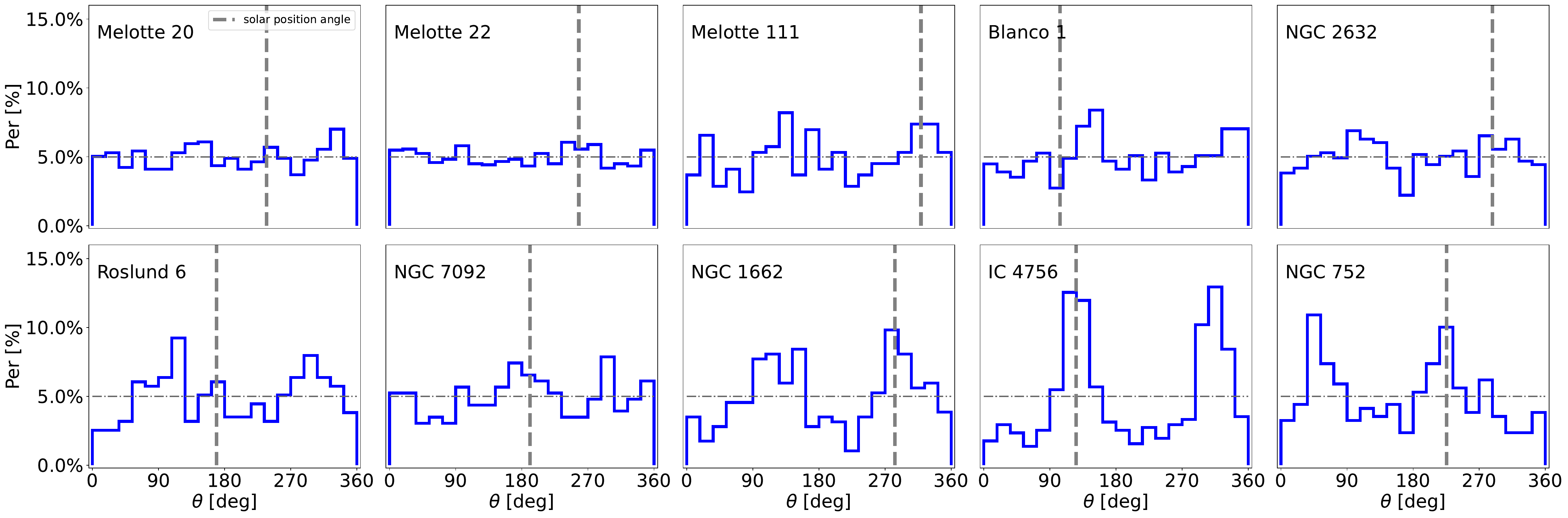}
  \caption{Distribution of the member stars as a function of the position angle $\theta$. The X-axis represents the position angle $\theta$ rotated counterclockwise around the projected motion [U$_{0}$, V$_{0}$] of the OCs. The solar position angle is indicated as the vertical gray dashed line. The 5\% level is shown as the horizontal dot-dashed line to help the illustration.}
  \label{fig:orbit-theta}
\end{figure*}

Figure~\ref{fig:orbit-theta} shows the fraction of cluster member stars as a function of position angle $\theta$, measured counterclockwise around each cluster's median velocity vector [U$_0$, V$_0$]. By design, $\theta=0^\circ$ (equivalently $\theta=360^\circ$) corresponds to the leading direction of the cluster's motion, whereas $\theta=180^\circ$ signifies the trailing direction. This figure illustrates if the member stars are isotopically distributed in angle. 

In the simplified picture of tidal-stripping processes, one might expect that a strongly perturbed star cluster could show pronounced stellar overdensities in the forward ($\theta=0^\circ$) and rearward ($\theta=180^\circ$) directions \citep[see e.g. the simulated and observed star cluster tidal tails in ][]{Jerabkova2021, Kroupa2022}. These correspond, respectively, to a ``leading tail'' and a ``trailing tail'' of stripped stars. However, from a straightforward inspection of Figure~\ref{fig:orbit-theta} none of the golden sample OCs shows a definitive peak at $\theta=180^\circ$ or $\theta=0^\circ$. In fact, for several cases of our golden sample OCs show no overdensity as a function of the position angle $\theta$, for instance, Melotte~20 and Melotte~22, indicating that they do not present any clear tails around the cluster. For some other cases (e.g., NGC~1662, IC~4756, NGC~752, Roslund~6), density peaks appear near $\theta \approx 90^\circ$ or $\theta \approx 270^\circ$, rather than at the leading or trailing directions, suggesting that at least near the center of the cluster, tidal features may lie perpendicular-rather than parallel-to the cluster's instantaneous velocity vector.

This result does not contradict the picture that tidal tails align with the orbital path. In fact, it is a support to numerical simulations of OCs and their tidal tail structures. For instance, in the tidal tail simulations of \citet{Jerabkova2021}, \citet{Kroupa2022} and \citet{Kroupa2024ApJ94K}, the extended long tidal tails ($\sim$ several hundred parsecs) do more or less follow the motion direction ($\theta=0^\circ$ or $\theta=180^\circ$ in our case) with an angle of the tidal tail long axis evolving with the cluster age as discussed extensively in \citet{Dinnbier2022}, while the innermost portion of the tidal tails often exhibits significant curvature as discussed by \citet{Kupper2010}. If one focuses on a relatively small field around the cluster center, i.e., extending only out to a few tidal radii, the inner portion of each tail can appear to sweep ``sideways'' in projection, creating a bend near the cluster center and appear as a near-perpendicular morphology before fanning out along the leading and trailing ($\sim 0^\circ /180^\circ$) directions at larger distances.

Observationally, it is very rare that one can capture every stripped star out to large distances from the cluster, due to the limiting magnitude constraint in photometry and tidal density compared to field stars. If the data is truncated, we primarily see the initial stages of the tidal arms, which may not yet be oriented parallel to the cluster's motion vector. Indeed, the tidal structure OC catalog we adopt from T22 selected member stars only 50 pc from the cluster center, and the angular density histogram of Figure~\ref{fig:orbit-theta} is built only the relatively inner tidal part around the cluster center. 

A sufficiently large survey area along the $\sim 0^\circ/180^\circ$ direction of our sample OCs may result in more extended tidal tails, especially for those clusters showing two density peaks near-perpendicular. Figure~\ref{fig:orbit-theta} can serve as a finding chart for such studies in the future.

\section{Summary}
\label{sec:sum}

As a well-defined single stellar population, OC serves as an ideal laboratory for studying the interplay between internal and external gravitational forces and the long-term evolution of star clusters.

In this study, we investigate the 3D structures of nearby OCs with tidal features based on the \citet{Tarricq2022} OC member catalog with tidal features. It is a catalog using high-precision astrometric data from the Gaia EDR3. While tidal structures in OCs have been extensively studied using 2D projections, analyzing their 3D morphology is essential to better understand their intrinsic structure and dynamical evolution. However, one of the major challenges in 3D studies of OCs arises from the systematic distortion effects introduced by parallax uncertainties, which can artificially elongate clusters along the line of sight. To address this issue, we develop a new method based on three-dimensional principal component analysis to identify a ``golden sample'' of OCs that are minimally affected by such distortions, allowing for more reliable 3D structure analysis.

We limit our selection to clusters within 500 pc and member stars more than 200, then a parallax zero-point correction is applied to refine distance estimates for all member stars (see Sec.~\ref{sec:plx}). Using PCA on the Galactocentric Cartesian coordinates of member stars, we measure the alignment between the 3D PCA first principal axis of each cluster and the observer's line of sight (see Sec.~\ref{sec:3dpca}). Clusters with a misalignment angle $\phi > 30^\circ$ are selected to ensure minimal elongation bias. The final golden sample consists of eight OCs: Blanco~1, Melotte~20, Melotte~22, NGC~2632, NGC~7092, NGC~1662, Roslund~6 and Melotte~111. These span ages from 51.3 Myr to 810 Myr, offering a diverse sample for probing dynamical evolution.

To characterize the structure of these clusters, we fit both 2D and 3D King profiles to their stellar density profiles. Our analysis reveals that, in most cases, the tidal radii derived from 3D modeling ($R_{t,{\rm 3D}}$) are systematically larger than those from 2D analysis ($R_{t,{\rm 2D}}$). This discrepancy may be because 2D projections tend to compress the apparent distribution of stars, leading to an underestimation of the tidal extent of the cluster. 

We suggest using the King profile deviation at the tidal radius to characterize the cluster structure asymmetry. The 3D King model reveals greater deviations ($\Delta \rho_{\rm 3D}$) from observed densities at the tidal radius than 2D analyses ($\Delta \rho_{\rm 2D}$), indicating that 3D fitting better captures tidal distortions.

Beyond structural parameters, we explore the relationship between cluster morphology and motion in the Galactic plane. By analyzing the spatial distribution of cluster members relative to their bulk motion, we investigate whether OCs exhibit preferential elongation along or perpendicular to their direction of motion. Simplified tidal theory predicts that clusters should develop leading and trailing tidal tails aligned with their orbital trajectory due to differential gravitational forces from the Milky Way. More recent simulations and observations reveal a more complex picture \citep{Jerabkova2021,Kroupa2022,Dinnbier2022}. Our analysis of the angular distribution of member stars supports this complex picture. While some clusters exhibit no preference for position angle, others display density peaks at perpendicular angles ($\sim 90^\circ$ or $\sim 270^\circ$). This feature can be interpreted as the inner part of more extended leading/trailing tails at larger distances. The current member star catalog may only detect the initial phases of tidal tail development. Future extended surveys covering larger areas have a great potential to trace tails fully aligned with orbital motion.

This work highlights the importance of considering 3D morphology when interpreting the tidal structures of OCs. The traditional approach of using 2D projections can lead to biased conclusions regarding cluster size, shape, and dynamical state. The golden sample clusters identified in this work can serve as a benchmark for future studies of tidal interactions and their impact on cluster evolution.

\begin{acknowledgements}

We thank the anonymous referee for the kind and helpful comments, which helped to improve the paper's clarity.

We thank Dr. Yoann Tarricq and Dr. Xiaoying Pang for their useful discussions. 

We acknowledge National Natural Science Foundation of China (NSFC) No. 12003001, No. 12203100, No. 12303026, No. 12203099, the China Manned Space Project with No. CMS-CSST-2021-A08. C.Y. acknowledges the Natural Science Research Project of Anhui Educational Committee No. 2024AH050049 and the Anhui Project (Z010118169). L.L. thanks the support of the Young Data Scientist Project of the National Astronomical Data Center. M.F. acknowledges the National Key Research and Development Program of China (No. 2023YFA1608100). H.Z. is funded by China Postdoctoral Science Foundation (No. 2022M723373), and the Jiangsu Funding Program for Excellent Postdoctoral Talent.

This work has made use of data from the European Space Agency (ESA) mission Gaia \url{(http://www.cosmos.esa.int/gaia)}, processed by the Gaia Data Processing and Analysis Consortium (DPAC, \url{http://www.cosmos.esa.int/web/gaia/dpac/consortium}). This research has made use of the TOPCAT catalog handling and plotting tool~\citep{topcat2}; the Simbad database and the VizieR catalog access tool, CDS, Strasbourg, France; and NASA's Astrophysics Data System.
\end{acknowledgements}

\bibliographystyle{aa.bst}
\bibliography{aa54212-25.bib}

\begin{appendix}
\section{The structure parameters of OCs}

 \begin{sidewaystable}
   \caption{The structure parameters of OCs.}
    \label{tab:structure-parameters}
    \centering
    \footnotesize 
    \begin{tabularx}{\textheight}{l|rr|rrrrr|rrrrr} 
        \hline
        & \multicolumn{2}{c|}{T22 (2D)} & \multicolumn{5}{c|}{2D this work} & \multicolumn{5}{c}{3D this work} \\
        \hline
        \multicolumn{1}{c|}{Cluster} &\multicolumn{1}{c}{$R_{\rm c}$} & \multicolumn{1}{c|}{$R_{\rm t}$} &\multicolumn{1}{c}{$k$} &\multicolumn{1}{c}{$R_{\rm c}$} &\multicolumn{1}{c}{$R_{\rm t}$} &\multicolumn{1}{c}{$c$} &\multicolumn{1}{c|}{$\Delta \rho_{\rm 2D}$} &\multicolumn{1}{c}{$k$} &\multicolumn{1}{c}{$R_{\rm c}$} &\multicolumn{1}{c}{$R_{\rm t}$} &\multicolumn{1}{c}{$c$} &\multicolumn{1}{c}{$\Delta \rho_{\rm 3D}$} \\
        &\multicolumn{1}{c}{(pc)} & \multicolumn{1}{c|}{(pc)} &\multicolumn{1}{c}{(pc$^{-2}$)} &\multicolumn{1}{c}{(pc)} &\multicolumn{1}{c}{(pc)} &\multicolumn{1}{c}{(pc$^{-2}$)} &\multicolumn{1}{c|}{($c$)} &\multicolumn{1}{c}{(pc$^{-3}$)} &\multicolumn{1}{c}{(pc)} &\multicolumn{1}{c}{(pc)} &\multicolumn{1}{c}{(10$^{-5}$ pc$^{-3}$)} &\multicolumn{1}{c}{($c$)} \\ 
        \hline
        Melotte 20&$3.64\pm0.34$&$22.10\pm1.48$&$11.26\pm0.91$&$3.27\pm0.28$&$25.99\pm1.74$&$0.012\pm0.001$& 1.194&$1.34\pm0.20$&$2.73\pm0.29$&$24.38\pm0.82$&$10.859\pm1.399$& 6.038 \\
        Melotte 22&$1.91\pm0.14$&$19.50\pm0.92$&$50.96\pm3.51$&$1.86\pm0.11$&$19.25\pm0.75$&$0.007\pm0.001$& 1.282&$14.50\pm2.62$&$1.06\pm0.11$&$20.41\pm0.47$&$4.925\pm0.841$& 12.289 \\
        Melotte 111&$2.04\pm0.62$&$19.02\pm5.03$&$0.87\pm0.18$&$7.72\pm1.90$&$29.95\pm5.79$&$0.011\pm0.001$& -0.051&$1.01\pm0.46$&$1.31\pm0.36$&$22.11\pm1.84$&$12.349\pm1.412$ & 2.973\\
        Blanco 1&$2.28\pm0.22$&$13.17\pm1.02$&$16.56\pm1.83$&$2.00\pm0.23$&$16.54\pm1.50$&$0.016\pm0.001$& 0.612 &$2.33\pm0.48$&$1.72\pm0.24$&$18.37\pm0.88$&$1.384\pm0.326$ & 99.018\\
        NGC 2632&$2.76\pm0.29$&$15.68\pm0.76$&$26.32\pm1.80$&$2.63\pm0.21$&$16.22\pm0.73$&$0.008\pm0.001$& 1.252&$4.97\pm0.73$&$1.80\pm0.19$&$16.74\pm0.46$&$6.967\pm1.017$ & 24.324\\
        Roslund 6&$4.28\pm0.63$&$32.11\pm5.32$&$2.36\pm0.33$&$4.17\pm0.61$&$37.55\pm5.43$&$0.007\pm0.001$& 0.354&$0.22\pm0.05$&$3.58\pm0.64$&$32.30\pm1.86$&$5.737\pm1.083$& 4.204 \\
        NGC 7092&$1.64\pm0.24$&$59.41\pm8.53$&$3.63\pm0.78$&$2.22\pm0.38$&$42.58\pm8.59$&$0.004\pm0.001$& 0.083&$1.18\pm0.79$&$0.99\pm0.36$&$42.41\pm3.08$&$2.405\pm0.612$& 0.857 \\
        NGC 1662&$1.56\pm0.18$&$38.32\pm8.27$&$6.37\pm1.41$&$1.77\pm0.30$&$38.88\pm7.60$&$0.008\pm0.001$& 0.202&$0.42\pm0.19$&$1.61\pm0.40$&$58.12\pm3.95$&$1.765\pm0.513$& 0.701 \\
        \hline
        IC 4756&$2.59\pm0.28$&$30.82\pm3.44$&$9.07\pm1.06$&$2.55\pm0.27$&$30.08\pm2.84$&$0.008\pm0.001$& -0.065&$0.75\pm0.24$&$1.64\pm0.29$&$52.92\pm2.43$&$0.812\pm0.151$& 7.115 \\
        NGC 752&$1.85\pm0.20$&$28.17\pm7.46$&$6.86\pm1.13$&$1.97\pm0.28$&$29.23\pm4.85$&$0.012\pm0.001$& -0.008&$0.41\pm0.20$&$1.52\pm0.40$&$71.02\pm4.09$&$0.098\pm0.031$& 10.942 \\
        \hline
    \end{tabularx}
    \tablefoot{Column ``Cluster" lists the OC names. All the other columns are parameters of the King model. $k$ is a proportionality constant related to the center density of the cluster, $R_c$ is the core radius of the cluster, $R_t$ is the tidal radius of the cluster. $c$ is a constant with units consistent with $k$. $\Delta \rho_{\rm 2D}$ represent the discrepancy between the observed 2D number density and the theoretical 2D number density at the tidal radius. $\Delta \rho_{\rm 3D}$ represent the discrepancy between the observed 3D number density and the theoretical 3D number density at the tidal radius.}
\end{sidewaystable}

\end{appendix}

\label{LastPage}
\end{document}